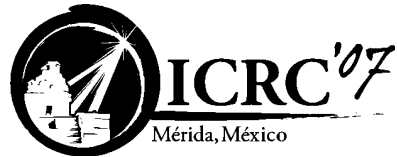

# Geant4 applications in the heliospheric radiation environment

PEDRO BROGUEIRA[1], PATRÍCIA GONÇALVES[2], ANA KEATING[2], DALMIRO MAIA[3], MÁRIO PIMENTA[2], BERNARDO TOMÉ[2]
[1]*IST, Instituto Superior Técnico*
[2]*LIP, Laboratório de Instrumentação e Física Experimental de Partículas*
[3]*FCUP, Faculdade de Ciências da Universidade do Porto*
bernardo@lip.pt

**Abstract:** The high energy ionizing radiation environment in the solar system consists of three main sources: the radiation belts, galactic cosmic rays and solar energetic particles. Geant4 is a Monte Carlo radiation transport simulation toolkit, with applications in areas as high energy physics, nuclear physics, astrophysics or medical physics research. In this poster, Geant4 applications to model and study the effects of the heliospheric radiation environment are presented. Specific applications are being developed to study the effect of the radiation environment on detector components, to describe the response and to optimise the design of radiation monitors for future space missions and to predict the radiation environment in Mars surface, orbits and moons.

## GEANT4 toolkit

Geant4 [1,2] is a Monte Carlo radiation transport simulation toolkit, with applications in areas as high energy physics, nuclear physics, astrophysics or medical physics research. It follows an Object-Oriented design which allows for the development of flexible simulation applications. Geant4 includes an extensive set of electromagnetic, hadronic and optics physics processes and tracking capabilities in 3D geometries of arbitrary complexity. The electromagnetic physics category covers the energy range from 250 eV to 10 TeV (up to 1000 PeV for muons) while hadronic physics models span over 15 orders of magnitude in energy, starting from neutron thermal energies. The optical physics process category allows the simulation of scintillation, Cherenkov or transition radiation based detectors. A distinct class of particles, the optical photons, is associated to this process category. The tracking of optical photons includes refraction and reflection at medium boundaries, Rayleigh scattering and bulk absorption.
The optical properties of a medium, such as refractive index, absorption length and reflectivity coefficients, can be expressed as functions of the photon's wavelength. The characteristics of the interfaces between different media can be defined using the UNIFIED optical model [2]. Full characterisation of scintillators include emission spectra, light yields, fast and slow scintillation components and associated decay time constants.

## Integrated radiation environment, effects and component degradation simulation tool.

The MarsREC Radiation Environment Module employs the Geant4 particle transport tool and includes parameters such as Martian time, detection position, solar longitude, solar cycle modulated cosmic ray and solar particle event spectra, 4-D EMCD atmosphere, geology and MOLA [4] topology.
The geometry implemented in Geant4 program takes into account:

- The pixel size given by the 5°x5° accuracy of the Mars Climate Database

- (MCD) [4], for each (long, latitude) location
- Average composition of the soil: 30% $Fe_2O_3$, 70% $SiO_2$, with 3.75 g/cm3 density;
- Thickness of the 32 atmospheric layers given by the sigma levels of the MCD;
- Fixed atmospheric composition consisting of 95% $CO_2$, 2.5% $N_2$, 1.25% Ar, 1.15% $O_2$, 0.07% CO and 0.03% $H_2O$.
- Atmospheric density, temperature and pressure given by the 32 layers of the atmospheric table computed from the MCD.
- Different times of the Martian Day correspond to different geometry set-ups

The framework developed is capable of:

- Predicting the high-energy radiation environment at the surface of Mars for different locations and solar longitudes.
- Tracking all primary and secondary particles, showing the relative importance of the backscattered neutron component of the radiation environment.
- Predicting radiation environment variation with climate changes along the Martian year.
- Evaluating Dose Equivalents and Dose depositions at the surface of Mars
- Calculating the energy spectra and particle species, radiation fluxes at component level, energy depositions and doses
- Computing SEU rates in specific components.

The results presented show that the framework is capable of calculating the energy spectra and particle species at any location on Mars with a resolution of 5º×5º. This MarsREC module also shows the relative importance of the backscattered component of the radiation environment, illustrating the importance of the Mars' geology in the radiation environment characterization. The Martian atmosphere, with a very low density (of at most $10^{-2}$ kgm$^{-3}$), behaves as a soft attenuator for incoming radiation capable of reaching the Mars surface, resulting in an important contribution from secondary particles generated and backscattered at the surface.

The dust density is typically less than $10^{-3}$ g/cm$^2$ which means less than $0.5 \times 10^{-3}$% of the atmospheric density. For this reason the impact of different dust scenarios is not expected to be very significant.

Fluences detected at the surface of Mars, due to the proton component in Galactic Cosmic Rays (GCR), are of the order of magnitude of $10^7$ to $10^8$ particles/cm$^2$.

The MarsREC Radiation Environment module was used as well to compute transfer functions and determine the effects of annual atmospheric variations on the radiation environment at the surface of Mars. Figure 1 illustrates the total fluence of particles at the surface of Mars as a function of solar longitude. This framework will be very valuable for planning future Mars missions and estimating the effects on lander systems, as well to predict the expected instrument behaviour.

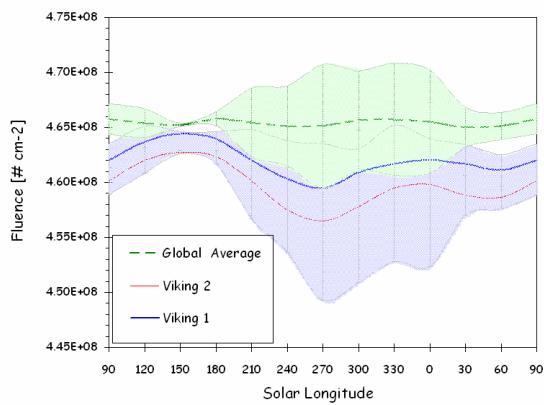

Figure 1: Total integrated fluence as function of Solar Longitude for different locations

Results show that while the total ionising doses at the surface of Mars are of lesser concern to EEE components, dose equivalents are of major concern for manned missions. Moreover the relative abundance of protons and neutrons may result in Displacement Damage and Single Event Effects.

MarsREC predictions were compared with experimental data and other software predictions. Results were in very good agreement.

## Simulation of a generic space radiation monitor concept

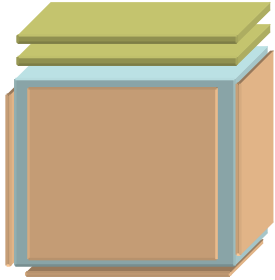

Figure 2: View of the simulated particle detector. From top to bottom: the two silicon planes and the scintillating crystal surrounded by 5 photodetectors.

The simulation of a simple space radiation monitor concept [5] was developed using the Geant4 toolkit. This concept, based on a scintillating crystal, was studied, consisting of an instrument capable of performing not only as a radiation switch, providing ancillary trigger information for the spacecraft, but also as a scientific instrument, which will measure fluxes and energy distributions of electrons, protons and ions in relevant energy ranges: 0.5-150 MeV per nucleon for protons and ions and 0.1-20 MeV for electrons.

The detector consists of a tracker made of two position sensitive silicon planes followed by a CsI(Tl) scintillating crystal surrounded by photodetectors (Figure 2).

In the present simulation the silicon tracker planes were 500 μm thick, the crystal size was 3x3x3 cm$^3$ thick. The scintillation properties of the CsI(Tl) crystal were used. These consisted of a light yield of 65000 photons per MeV of deposited energy and two scintillation components with decay time constants of 0.68 μs (64%) and 3.34 μs (36%). The scintillation light is readout by large area silicon PIN photodiodes, with a spectral sensitivity matching the CsI(Tl) emission spectrum. Since the silicon photodiodes are sensitive both to photons and charged particles, this makes them suitable to be used also as anti-coincidence shield. Particle identification is performed by measuring the energy loss in the thin silicon trackers.

Figure 3 shows the simulated energy loss in the first (top) and second (bottom) silicon planes, as a function of the initial particle energy, for electrons, protons and alphas. Figure 4 shows the energy deposited in the crystal by electrons, protons and alpha particles as a function of their kinetic energy before entering the detector.

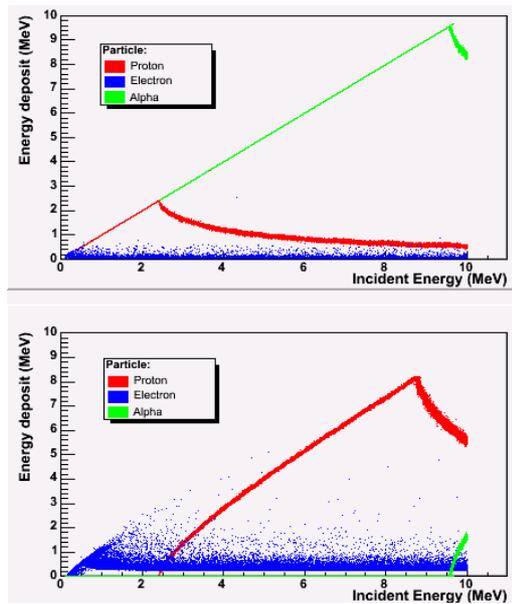

Figure 3: Energy deposited in the first (top) and second (bottom) tracker planes by electrons (blue dots), protons (red dots) and alpha particles (green dots), as a function of the energy of the incoming particle.

The instrument geometry and corresponding structural model implemented within the framework of the Geant4 simulation toolkit were used to perform several studies, such as:

- Study of the interaction of secondary particles induced by the structural frame and electronics circuits.
- Generation of the response function for the instrument simulator.
- Study of the radiation dose to be expected by electronics, calorimeters and trackers.
- Definition of typical observation scenarios and the related input spectra for

electrons, alpha particles and protons at different times.
- Device performance assessment as a radiation monitor and as a scientific instrument.

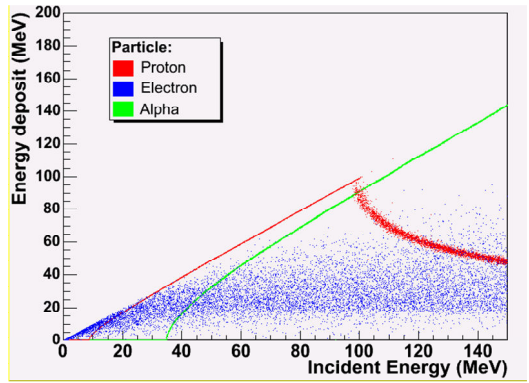

Figure 4: Energy deposited in the crystal by electrons (blue dots), protons (red dots) and alpha particles (green dots), as a function of the energy of the incoming particle.